\documentclass[preprint2]{aastex631}

\newcommand{\kms}          {\mbox{${\rm km~s^{-1}}$}}

\newcommand{\e}            {\mbox{$^{-1}$}}

\def\cm2{\mbox{${\rm cm^{-2}}$}}
\def\h2{\mbox{${\rm H}_2$}}
\def\nh2{\mbox{$n_{\rm H_2}$}}
\def\Nh2{\mbox{$N_{{\rm H}_2}$}}
\def\Mh2{\mbox{$M_{{\rm H}_2}$}}

\def\simgt{\lower.5ex\hbox{$\; \buildrel > \over \sim \;$}}
\def\simlt{\lower.5ex\hbox{$\; \buildrel < \over \sim \;$}}

\def\13CO{$^{13}$CO}
\def\C18O{C$^{18}$O}
\def\H2{H$_2$}

\def\Mdisk{\mbox{$M_{\rm disk}$}}

\def\vdrift{\mbox{$v_{\rm d}$}}
\def\tdrift{\mbox{$t_{\rm d}$}}
\def\RCO{\mbox{$R_{\rm CO}$}}
\def\RCOninety{\mbox{$R_{\rm CO,90\%}$}}
\def\Rdust{\mbox{$R_{\rm dust}$}}
\def\Rdustninety{\mbox{$R_{\rm dust,90\%}$}}
\def\Rc{\mbox{$R_{\rm c}$}}
\def\Qmin{\mbox{$Q_{\rm min}$}}
\def\Sigmagas{\mbox{$\Sigma_{\rm gas}$}}

\def\startfigcap{\vspace*{2.0\baselineskip}\bgroup\leftskip 0.45in\rightskip 0.45in\small} \def\endfigcap{\par\egroup\vspace*{2.0\baselineskip}}

\def\startfigcapside{\vspace*{2.0\baselineskip}\bgroup\leftskip 4.0in\rightskip 0.1in\small}
\def\endfigcapside{\par\egroup\vspace*{2.0\baselineskip}}

\usepackage{color}
\definecolor{CornflowerBlue}{rgb}{0.39,0.58,0.93}
\definecolor{TitleBrown}{rgb}{0.60,0.55,0.52}   
\definecolor{SectionRed}{rgb}{0.55,0.15, 0.17}
\definecolor{royalblue}{rgb}{0.25,0.41,0.88}
\definecolor{pureblue}{rgb}{0.0,0.0,1.0}
\definecolor{darkblue}{rgb}{0.05, 0.05, 0.45}
\definecolor{firebrick1}{rgb}{1.00,0.19,0.19}

\usepackage{amsmath}
\received{August 7, 2024}
\accepted{\today}
\submitjournal{ApJ}

\shorttitle{Dust drift timescales in protoplanetary disks}
\shortauthors{Williams et al.}

\begin{document}

\title{Dust Drift Timescales in Protoplanetary Disks at the Cusp of Gravitational Instability}

\correspondingauthor{Jonathan Williams}
\email{jw@hawaii.edu}

\author[0000-0001-5058-695X]{Jonathan P. Williams}
\affiliation{Institute for Astronomy, University of Hawai'i at Mānoa, 2680 Woodlawn Drive, Honolulu, HI 96822, USA}

\author[0009-0003-6274-657X]{Caleb Painter}
\affiliation{Institute for Astronomy, University of Hawai'i at Mānoa, 2680 Woodlawn Drive, Honolulu, HI 96822, USA}
\affiliation{Department of Astronomy, Harvard University $|$ Cambridge, MA 02138}

\author[0000-0002-4876-630X]{Alexa R. Anderson}
\affiliation{Institute for Astronomy, University of Hawai'i at Mānoa, 2680 Woodlawn Drive, Honolulu, HI 96822, USA}

\author[0000-0003-3133-3580]{Alvaro Ribas}
\affiliation{Institute of Astronomy, University of Cambridge, Madingley Road, Cambridge CB3 0HA, UK}

\begin{abstract}
Millimeter emitting dust grains have sizes that make them susceptible to drift in protoplanetary disks due to a difference between their orbital speed and that of the gas. The characteristic drift timescale depends on the surface density of the gas. By comparing disk radii measurements from ALMA CO and continuum observations at millimeter wavelengths, the gas surface density profile and dust drift time can be self-consistently determined. We find that profiles which match the measured dust mass have very short drift timescales, an order of magnitude or more shorter than the stellar age, whereas profiles for disks that are on the cusp of gravitational instability, defined via the minimum value of the Toomre parameter, $\Qmin\sim1-2$, have drift timescales comparable to the stellar lifetime. This holds for disks with masses of dust $\gtrsim 5\,M_\oplus$ across a range of absolute ages from less than 1\,Myr to over 10\,Myr. The inferred disk masses scale with stellar mass as $\Mdisk\approx M_\ast/5\Qmin$. This interpretation of the gas and dust disk sizes simultaneously solves two long standing issues regarding the dust lifetime and exoplanet mass budget and suggests that we consider millimeter wavelength observations as a window into an underlying population of particles with a wide size distribution in secular evolution with a massive planetesimal disk.
\end{abstract}


\section{Introduction}
\label{sec:introduction}
An important, or even defining, characteristic of planet-forming disks is that they contain dust grains that have grown by multiple orders of magnitude from the sub-micron sizes in the interstellar medium (ISM). With proportionally more mass per surface area, the effect of gas pressure is smaller for these large grains and they dynamically decouple from the gas, falling down to the midplane and drifting inwards towards the central star. The separation of solids from gas is the first step toward planetesimals and the assembly of planets through core accretion \citep{2010AREPS..38..493C}.

Millimeter wavelength observations of disks reveal the large dust grains in disks that are absent in the ISM and provide a window into the planet formation process.
The interpretation of the data contain two persistent puzzles, however.
First, the separation of solids from gas should lead to very rapid loss of the millimeter emitting dust, on timescales much shorter than observed disk lifetimes \citep{2007A&A...469.1169B}.
Disk gas sizes are typically about three times that of the millimeter-emitting dust which is greater than expected from their differing optical depths \citep{2019A&A...629A..79T}, demonstrating that dust has indeed drifted inwards but then stabilised at a finite radius. There are no identifiable trends in the gas-to-dust size ratio with stellar mass, disk dust mass, or substructure \citep{2022ApJ...931....6L}.
Second, dust masses are low, typically much less than the minimum mass of solids required to form the Solar System, $30\,M_\oplus$ \citep[e.g.,][]{2016ApJ...828...46A, 2016ApJ...831..125P} and the known exoplanet population \citep{2018A&A...618L...3M}. This implies either very rapid planet formation, late-stage infall onto disks around optically visible stars \citep{2023A&A...670L...8G}, or a reservoir of material that is unseen in the observations possibly due to high optical depth \citep{2019ApJ...877L..18Z, 2020A&A...642A.171R}.
The question remains of what exactly are the millimeter emitting grains in disks telling us about planet formation? 

The aerodynamics of dust grains in a gas disk is governed by the Stokes number, St, which is the ratio of the stopping time due to gas drag to the orbital period.
Relatively large, massive particles have long stopping times, high Stokes numbers, and are essentially unaffected by gas dynamics. On the opposite end of the scale are small, low mass, grains with low Stokes numbers that are swept along with the gas.
At intermediate values, St\,$\sim 1$, particles lose momentum on orbital timescales and move inwards at a speed, $\vdrift = [2{\rm St}/(1+{\rm St}^2)]v_0$, where
\begin{equation}
v_0=c_{\rm s}^2/2v_{\rm K}
\label{eq:v0}
\end{equation}
is the maximum drift speed and $c_{\rm s}, v_{\rm K}$ are the sound speed and Keplerian rotation speed respectively \citep{2010AREPS..38..493C}. Typical values are $v_0\simeq 10^{-2}\,\kms\ = 2\times 10^{-3}\,{\rm au}\,{\rm yr}^{-1}$ which is fast compared to $\sim 10^2$\,au disk size scales and Myr lifetimes, and may explain why many pre-main sequence stars even in very young regions are diskless \citep{2010ApJS..186..111L}.
Substructures that reverse local pressure gradients and thereby reduce gas-dust velocity differentials are an oft-quoted solution to this dust lifetime problem \citep{2012A&A...538A.114P}. This is clearly seen in transition disks with large central dust cavities \citep{2020ApJ...892..111F} and indeed most disks show substructures at varying levels of contrast \citep{2020ARA&A..58..483A}.
Alternatively, the particles that drift the most may be larger than those detected in infrared and millimeter wavelength observations. As long as the particles are smaller than the mean free path of the gas molecules, the conditions for Epstein drag apply and the St\,$=1$ particle size scales linearly with the gas surface density. Thus massive disks can sustain a population of larger dust grains at larger radii than lower mass disks.

Continuum observations show that disk sizes are smaller at longer wavelengths, showing that large grains drift inwards further than small grains.
\citet{2017ApJ...840...93P} proposed the idea that the disk radius-grain size relation can be used to infer the gas surface density and thereby constrain the disk gas mass, a fundamental but notoriously hard-to-measure quantity \citep{2017ASSL..445....1B}. Under the assumption that the Stokes number falls below unity at the observed edge of the dust, $\Rdust$, \citet{2019ApJ...878..116P} showed that the gas surface density at this location is
\begin{equation}
\Sigmagas(\Rdust) = \frac{2.5v_0\tdrift \rho_{\rm grain} s}{\Rdust}
\label{eq:Powell}
\end{equation}
where $s$ is the grain size, $\rho_{\rm grain}$ is the grain mass density, $\tdrift$ is the drift timescale,
and $v_0$ is defined in equation~\ref{eq:v0}.
A key step in their analysis is to set $\tdrift = t_\ast$, the stellar age, based on a picture where dust grain growth and radial drift are in equilibrium and have been occurring over the lifetime of the disk.
Then, using millimeter--centimeter wavelength observations, they constrain \Sigmagas\ at several radii and integrate over a fitted density profile to determine the masses of seven disks.

\citet{2022A&A...657A..74F} subsequently tested this method via disk synthesis models incorporating dust evolution. They confirmed that dust line locations depend on the gas surface density but are also inter-dependent at different grain sizes due to conservation of mass flux. The effect is most severe for large grains at small radii that are observed at long wavelengths. They show that the adopted drift time is a major source of uncertainty but suggested that the method could work if calibrated by other measures of disk structure.

In this paper, we re-examine the dust line idea but with a focus on measuring, rather than assuming a value for, the dust drift timescale based on a surface density constraint from CO observations. We show that drift times are short for the low disk masses inferred from standard conversions of the millimeter continuum flux but are comparable to stellar ages for the much higher gas densities in disks that are on the cusp of gravitational instability. In the latter case, we arrive at the same conclusion as \citet{2017ApJ...840...93P} but based on an observable metric of the disk size in CO rather than assuming a balance between grain growth and radial drift.
Furthermore, this interpretation provides a single, combined solution to the afore-mentioned disk mass and lifetime puzzles. We begin by describing the surface density constraints in \S\ref{sec:constraints}, model a collection of disks with resolved dust and gas sizes in \S\ref{sec:results}, discuss the results and implications in \S\ref{sec:discussion} and summarize our work in \S\ref{sec:summary}.

\section{Surface density constraints}
\label{sec:constraints}
The strongest lines from the molecular gas in the cool, outer regions of protoplanetary disks are rotational transitions of CO. The CO molecule is relatively abundant and has a simple, well understood chemistry. A key finding for this work is the result of \citet{2023ApJ...954...41T} who showed that the CO emitting disk extends out to a surface density threshold that depends only mildly on the disk mass and stellar luminosity and is largely immune to disk geometrical and structural parameters.
At a radius \RCOninety, defined as containing 90\% of the total flux of the CO $J=2-1$ line, the surface density is
\begin{equation}
\begin{split}
\Sigmagas&(\RCOninety) = \\
 & 10^{-2.15-0.53\log_{10}L_\ast}\left(\frac{M_{\rm gas}}{M_\odot}\right)^{0.3-0.08\log_{10}L_\ast}\ {\rm g\,cm^{-2}}.
\end{split}
\label{eq:Trapman}
\end{equation}
This useful result arises because the CO emission becomes optically thin at about the same column density where the molecule is photo-dissociated.
The mass and luminosity dependencies are due to their effect on the CO abundance through freeze-out.

Many disks have been mapped with the Atacama Large Millimeter Array (ALMA) at sufficient sensitivity and resolution to determine \RCOninety\ and the corresponding \Rdustninety\ at a neighboring continuum wavelength $\simeq 1.3$\,mm. The choice of 90\% for the radius definitions is somewhat arbitrary but is high enough that it is a good measure of the extent of each species as they tend to fall off rapidly as an exponential rather than a power law \citep{2020ARA&A..58..483A} yet not so high as to require very high signal-to-noise data. We explore the effect of different radii definitions in \S\ref{sec:robustness}.

The CO radius provides an anchor point for the gas surface density at the outer edge of the disk. The dust radius relates to the surface density about 3 times closer to the star but with a value that scales directly with \tdrift\ following equation~\ref{eq:Powell} where we use $\rho_{\rm grain}=2\,{\rm g\,cm}^{-3}$, and particle size, $s=\lambda/2\pi$ based on Mie theory.
For an accretion disk surface density profile,
\begin{equation}
\Sigmagas(R)\propto\left(\frac{R}{R_{\rm c}}\right)^{-\gamma}\exp\left[-\left(\frac{R}{R_{\rm c}}\right)^{2-\gamma}\right],
\label{eq:LyndenBellPringle}
\end{equation}
we can then solve for the characteristic radius, $\Rc$,
\begin{equation}
\Rc = \left\{\frac{\RCO^{2-\gamma} - \Rdust^{2-\gamma}}{\ln\left[\left(\frac{\Sigmagas(\Rdust)}{\Sigmagas(\RCO)}\right)\left(\frac{\Rdust}{\RCO}\right)^\gamma\right]}\right\}^{1/(2-\gamma)},
\label{eq:Rc}
\end{equation}
where we have dropped the 90\% suffix for legibility.
For a given $\gamma$, we therefore only require a normalization constraint to determine the gas surface density profile across the disk

We consider two cases that effectively represent the minimum and maximum possible disk masses:
\begin{enumerate}
\item the total disk mass is 100 times the measured dust mass;
\item the surface density is at the threshold for gravitational instability.
\end{enumerate}
For case 1, the dust mass is defined in the canonical way as $M_{\rm dust} = F_\nu d^2/\kappa_\nu B_\nu(T)$ where $F_\nu$ is the continuum flux density, $d$ is the distance, $\kappa_\nu=10\nu_{\rm GHz}\,{\rm cm}^2\,{\rm g}^{-1}$ is the dust grain opacity, $B_\nu(T)$ is the blackbody at the dust temperature $T=20$\,K, and we assume an ISM gas-to-dust ratio of 100 to determine the total disk mass \citep[e.g.,][]{2023ASPC..534..501M}.
For case 2, the gravitational stability is assessed through the Toomre parameter, $Q=c_{\rm s}\Omega/\pi G\Sigmagas$, which is required to be greater than one at all radii.
Here, $c_{\rm s}$ is the sound speed which we calculate using the same midplane temperature radial profile as in \citet{2017ApJ...840...93P}, and $\Omega=(GM_\ast/R^3)^{1/2}$ is the Keplerian rotation rate.
We require $Q$ to be greater than 1 at all radii to avoid rapid disk disintegration \citep{2001ApJ...553..174G}.

The surface density profile is constrained by its integrated value for case 1 and the minimum of the derived Toomre parameter for case 2. From the fitted value of \Rc, we are then able to determine \tdrift.
Due to the probable co-dependence of different dust lines shown by \citet{2022A&A...657A..74F}, we do not use multi-wavelength continuum data and only match the dust line location at a single wavelength, $\lambda\sim 1.3$\,mm, determined from ALMA Band 6 observations.
This traces relatively small grains where the drift position is largely dictated by the gas density.
The CO surface density constraint includes a disk mass dependence which we include based on a gas-to-dust ratio of 100 for the first case of the dust mass constraint and iterate to a solution for $\Sigmagas(R)$ for the second case of gravitational stability.

\section{Results}
\label{sec:results}
We looked through the literature for protoplanetary (Class II) disks with measured CO 2--1 and dust continuum sizes. \citet{2018ApJ...859...21A} provided a table of \RCOninety\ for a sample of Lupus disks and we extrapolated the fits to the ALMA Band 6 continuum visibilities in \citet{2021MNRAS.506.2804T} to determine \Rdustninety. \citet{2022ApJ...931....6L} looked at the ratio between gas and dust sizes to study the role of angular momentum transport in disk evolution and provided a useful table with \RCOninety\ and \Rdustninety\ for 26 sources. \citet{Semenov2024PRODIGE} fit power laws to IRAM observations to measure disk edges to determine an equivalent \RCOninety\ in four Taurus Class II sources and we scaled the dust radii measured at the 68\% cumulative threshold in \citet{2018ApJ...861...64T} to estimate \Rdustninety\ using an empirical scaling based on \citet{2022ApJ...931....6L}, $\Rdustninety = R_{\rm dust,68\%}/0.67$ (see \S\ref{sec:robustness}).
Finally we added the isolated, old and compact disk, MP Mus, using data from \citet{Ribas2023MPMus}.
The full sample amounts to 41 disks. The radii, dust mass, and other stellar properties used in this analysis are listed in Table~1 in the Appendix.

\begin{figure}
\begin{center}
\includegraphics[scale=0.42]{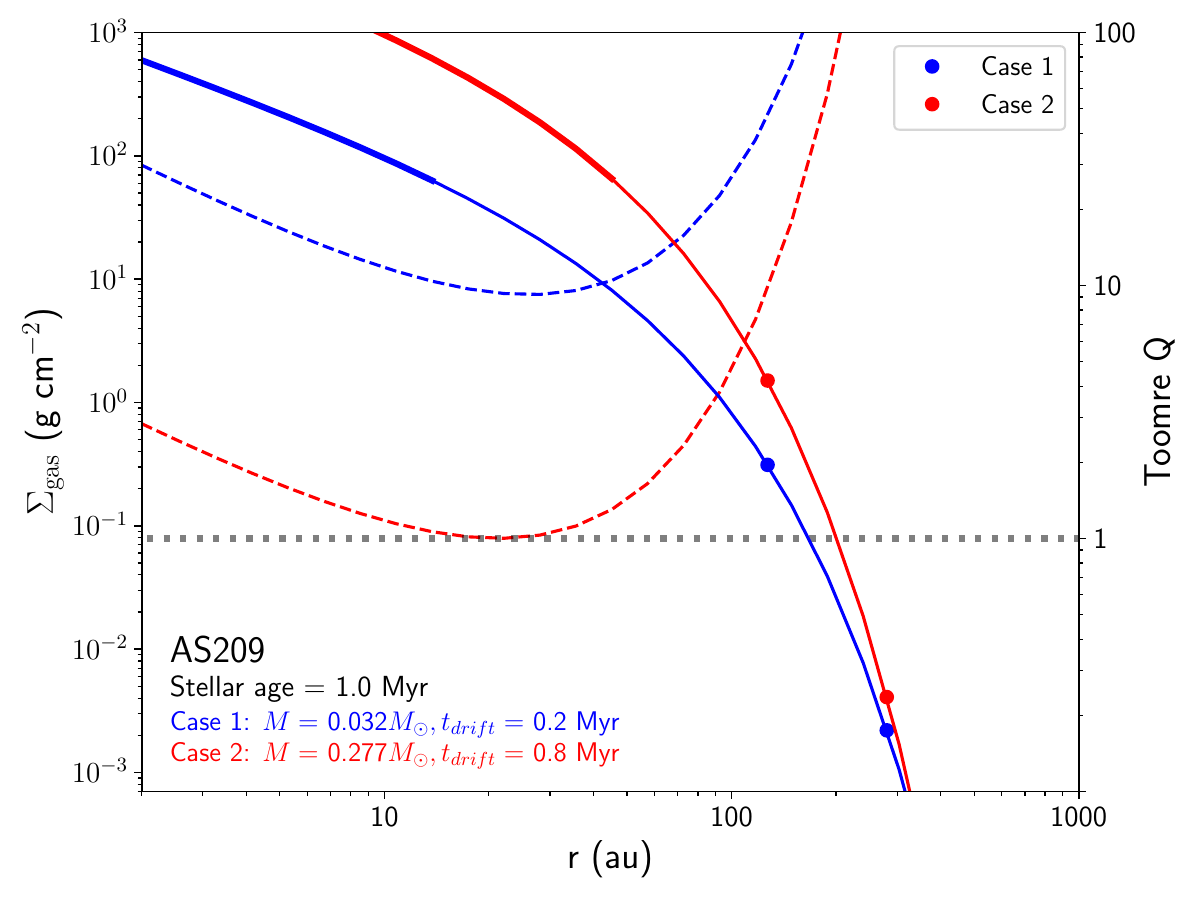}
\end{center}
\caption{Surface density fits to the CO and dust radius in the AS\,209 disk for the two cases of matching the dust mass at a gas-to-dust ratio of 100 (blue lines) and for a disk on the cusp of gravitational instability with $\Qmin=1$ (red lines). The points are at the observed dust and CO radii, measured at the 90\% cumulative flux threshold, with surface densities determined by equations~\ref{eq:Powell} and \ref{eq:Trapman} respectively. The solid lines show the surface density profiles (which are thicker at $\Sigmagas>50\,{\rm g\ cm}^{-2}$ where the 1.3mm continuum is optically thick), and the dashed lines the Toomre $Q$ parameter. The stellar age, disk masses, and inferred drift times, as derived by the scaling of the surface density at the dust radius, are given in the bottom left corner.}
\label{fig:AS209}
\end{figure}

Example fits to the surface density profile for AS\,209, the first source in the alphabetically ordered Table~\ref{tab:sample}, is shown in Figure~\ref{fig:AS209}.
Here we have fixed $\gamma=1$ but we discuss the effect of different values for this parameter in \S\ref{sec:robustness}.
The points show the gas surface densities at the \Rdustninety\ and \RCOninety\ locations, which differ from case 1 (blue) to case 2 (red) due to the drift time and mass dependence respectively.
The dust mass constraint provides a lower limit to the surface density and therefore drift time whereas the gravitational stability constraint sets an upper limit to the surface density and a maximum value for the drift time.

The drift time for the dust mass constraint is only 0.2\,Myr which is about 5 times shorter than the estimated age of the star \citep{2022ApJ...931....6L}. Some mechanism is therefore required to maintain the dust at large radii and indeed AS\,209 shows a set of narrow, highly confined dust rings \citep{2018ApJ...869L..48G}.
However, given the stellar mass and luminosity, the disk could be almost ten times more massive, $\sim 0.28\,M_\odot$, than inferred from the continuum flux and be on the margins of gravitational instability. The higher surface density in such a disk provides an alternative way to keep the dust at large radii. Notably, the inferred drift time in this case, 0.8\,Myr, is much more similar to the stellar age.

The fits for each of the other 40 disks are shown in Figures~\ref{fig:panelplot1},~\ref{fig:panelplot2} in the Appendix. As with AS\,209, we find that dust drift times are relatively short compared to the stellar age if the disk mass is 100 times the dust mass (typically $10^{-2}-10^{-3}\,M_\odot$ for this sample).
The results for the case of a marginally gravitationally stable disk is also similar to AS\,209 in the sense that drift times are much longer and tend to be similar to the stellar age. This holds over a wide range of disk masses, sizes (whether CO or dust), and whether the star is very young, $<1$\,Myr, or old, $>10$\,Myr. The implied disk masses for this case range from 0.05 to $0.67\,M_\odot$.

\begin{figure*}
\begin{center}
\includegraphics[scale=0.8]{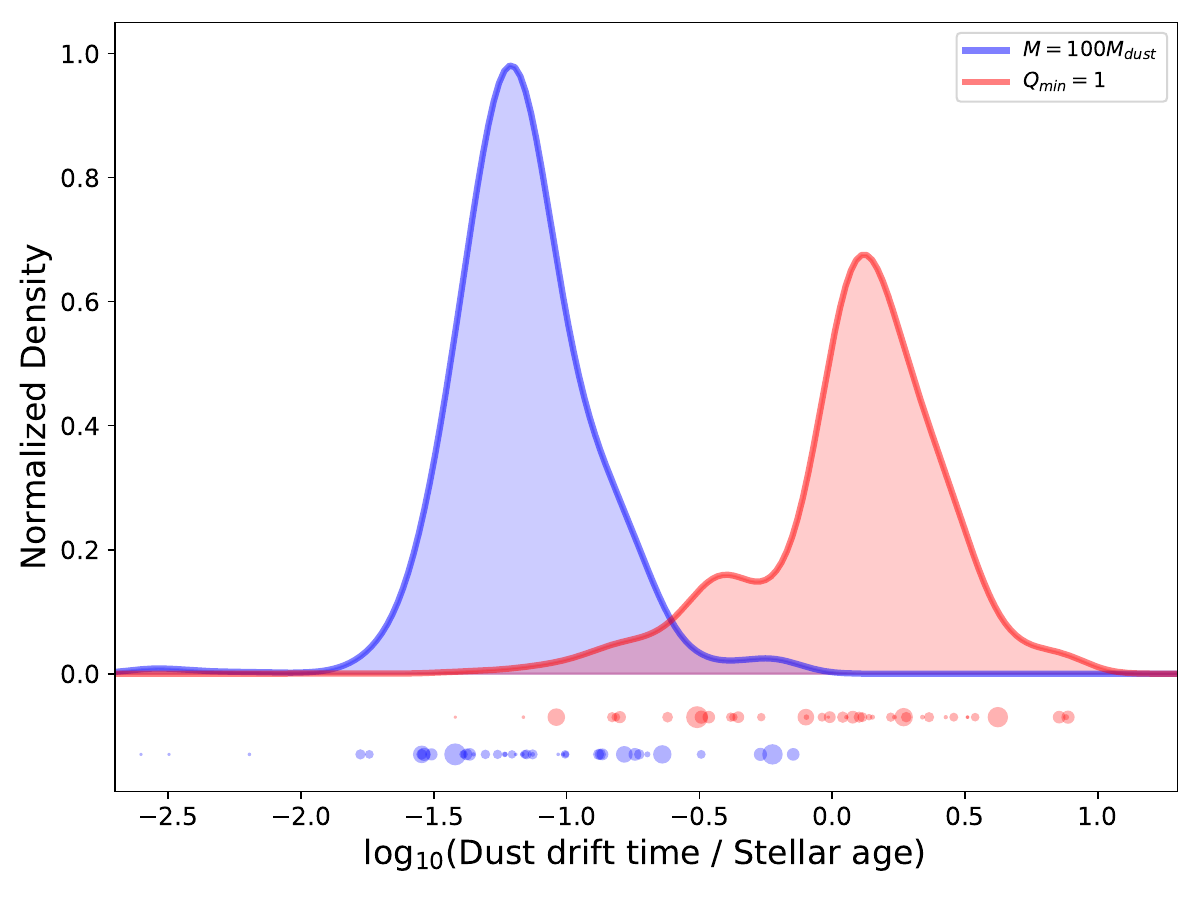}
\end{center}
\caption{Probability density functions of the ratio of drift time to stellar age for the 41 disks in our sample. The blue curve shows case 1 where the disk mass equals 100 times the dust mass and the red curve represents case 2 where a marginally gravitationally stable disk has Toomre $Q$ minimum equal to 1. The functions are determined through kernel density estimation of the points shown at the bottom of the plot, where the symbol size is proportional to the dust mass.}
\label{fig:drift_times}
\end{figure*}

The distributions of drift times relative to stellar ages for the two cases are shown in Figure~\ref{fig:drift_times}. The distributions are quite narrow and distinct from each other with medians, $\log_{10}(\tdrift/t_\ast) =-1.17\pm0.52$ (case 1) and $0.05\pm0.55$ (case 2).
There are a few outliers in each case. For the case 1 dust mass constraint, the three disks (FP\,Tau, J04202555+2700355, and the spectroscopic binary V4046\,Sgr) with very short relative drift times have the lowest dust masses in the sample, $\leq 5\,M_\oplus$. This may indicate that the disks are in the process of dispersal, but it is not a sufficient condition as there are other very low mass disks with longer drift times. The three disks with the largest $\tdrift/t_\ast$ ratios either show spiral structure and are very young, $t_\ast<0.5$\,Myr \citep[IM\,Lup, WaOph6;][]{2018ApJ...869L..41A} or are a hierarchical triple \citep[UZ\,Tau;][]{1996AJ....111.2431J} and thereby externally influenced.
For the case 2 Toomre stability constraint, the outliers at the low end are the same FP\,Tau and J04202555+2700355 but notably not V4046\,Sgr, which has a drift time slightly greater than its advanced stellar age of 25\,Myr. IM\,Lup, WaOph6, and UZ\,Tau are at the high ratio end again, and joined by Sz\,65 which is a compact disk in a wide binary \citep{2024A&A...682A..55M} and, like UZ\,Tau, possibly affected by external forces.

We confirm the known dust lifetime problem that millimeter emitting dust grains should rapidly drift inwards on timescales much shorter than the stellar age and show that this holds if the disks have relatively low masses as derived from the millimeter continuum luminosity and a gas-to-dust ratio of 100. However, our main result is that more massive disks, at the cusp of gravitational instability, have drift timescales within a factor of 3 of the stellar age.

\bigskip
\section{Discussion}
\label{sec:discussion}
The outer edge of the CO emission from a protoplanetary disk occurs at a well defined column density. Using this as an anchor, we can then determine the maximally stable surface density profile using the Toomre $Q$ parameter that depends weakly on the stellar mass through the Keplerian rotation rate and even more weakly on the stellar luminosity through the gas sound speed. We can then calculate the drift time at the outer edge of the dust disk using equation~\ref{eq:Powell} and find that they are comparable to the age of the disk-hosting star, over a wide range from less than 1 to over 10\,Myr. The similarity between these two independent quantities requires explanation.

\subsection{Robustness}
\label{sec:robustness}
\begin{figure*}
\begin{center}
\includegraphics[width=\textwidth]{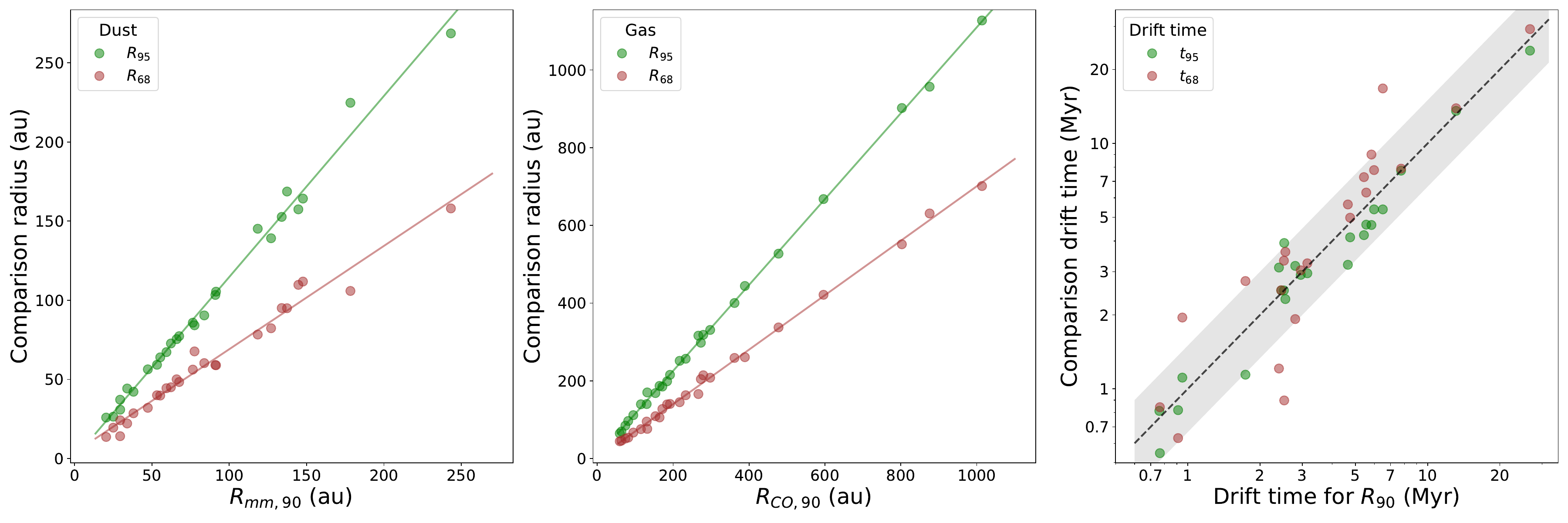}
\end{center}
\caption{Comparison of disk radii in the dust (left panel) and gas (middle panel), and case 2 drift times (right panel) for cumulative flux thresholds of 68\% and 95\% versus the fiducial 90\% threshold used in this work. The radii are tightly correlated in both dust and gas and drift times at the 90\% and 95\% levels agree to within 50\%.}
\label{fig:radius}
\end{figure*}

First, we checked that the result is robust to some of the implicit assumptions. The disk outer radii for both gas and dust were defined to be where 90\% of the CO or continuum flux, respectively, was contained.
\citet{2022ApJ...931....6L} provides a table of radii measurements at the 68\%, 90\% and 95\% cumulative flux threshold and we carried out the same analysis for each radius definition in this subsample of 26 disks.
Figure~\ref{fig:radius} shows the tight linear relations between the different radii with $R_{\rm dust,68\%}=0.67\Rdustninety$,  $R_{\rm dust,95\%}=1.15\Rdustninety$, and similarly in the gas, $R_{\rm CO,68\%}=0.71\RCOninety$,  $R_{\rm CO,95\%}=1.10\RCOninety$.
This translates to similar case 2 drift times across the full range in the sample, agreeing to within 50\% for the $R_{95\%}$ definition but more noisy for $R_{68\%}$.
This is not too surprising given that continuum radial profiles are typically sharply tapered at the outer edge and well traced by the 90\% and 95\% thresholds though less so at 68\% \citep{2020ARA&A..58..483A}. Nevertheless, given that there is no systematic bias in the drift times, the similarity to the stellar age is unchanged.
Much higher signal-to-noise ratios are required to measure disk radii at the 95\% cumulative flux threshold than at 90\% and the latter is sufficient to extend this work to a larger sample.

\begin{figure}
\begin{center}
\includegraphics[scale=0.42]{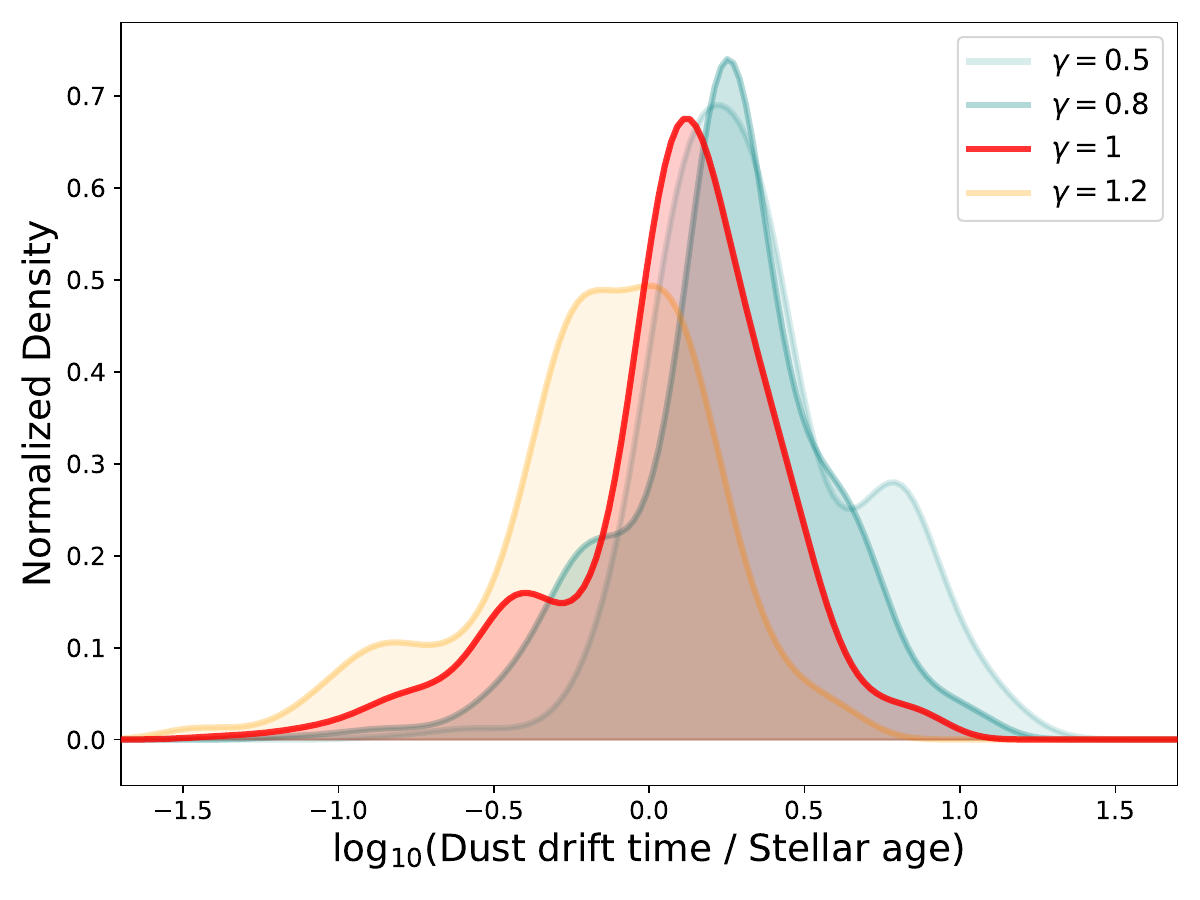}
\end{center}
\caption{The ratio of case 2 drift time to stellar age for different values of the power law index $\gamma$ used in the surface density fitting. Shallower profiles have lower central densities and longer drift times. As long as the gas density does not rise very strongly toward the center, $\gamma\leq 1.2$, then the result that the drift time for disks with $\Qmin=1$ is similar to the stellar age holds.}
\label{fig:gamma}
\end{figure}

The surface density profile in equation~\ref{eq:LyndenBellPringle} contains three parameters but we only have two constraints, $\Sigmagas(\RCO)$, and $\Qmin=1$. We fixed $\gamma=1$ initially but explore different values in Figure~\ref{fig:gamma} and find that the drift time is inversely related to $\gamma$ but the dependence is relatively slight and the median value of drift time to stellar age brackets unity for $\gamma=0.5$ to 1.2. Very steep density gradients, $\gamma=1.5$, generally do not fit the CO and dust radii as well but, when they do, imply shorter drift times with a mean, $\log_{10}(\tdrift/t_\ast) =-0.71\pm0.48$. The reason is that disks with such steep density profiles can become gravitationally unstable at small radii for lower disk masses with relatively low densities at the dust line location.
As long as surface density profiles have $\gamma\leq 1.2$, which is verified in a few detailed studies of well resolved disks \citep[e.g.,][]{2016ApJ...830...32W}, then the basic result that case 2 drift times are comparable to stellar ages holds.

The drift time is inversely proportional to the grain density, $\rho_{\rm grain}$, in equation~\ref{eq:Powell}. We have fixed the density at 2\,g\,cm$^{-2}$ which is intermediate between water ice, organics and silicates and similar to commonly used values in the dust dynamics literature \citep{2018ApJ...869L..45B}. A composition rich in silicates or iron rich minerals could have higher densities but we expect a large mass fraction in ices and organics at the cold temperatures for the large radii under consideration here so this density is a reasonable upper limit, thus setting a lower limit to \tdrift. In fact, the density could be lower due to grain porosity, which is a significant unknown \citep{2023arXiv231213287B}, and \tdrift\ would then be correspondingly higher.

The stellar age measurements also add some uncertainty. We use the values from the same literature sources for the gas radii measurements \citep{2022ApJ...931....6L, Semenov2024PRODIGE, Ribas2023MPMus} which are in turn based on pre-main sequence evolutionary tracks, most commonly Baraffe and MIST models. There are persistent differences between models but these are relatively small, $\sim 0.5$\,Myr, compared to the range of ages in our sample. This could account for some of the dispersion in the drift time to stellar age ratio but it is not sufficient to significantly change the mean value.

\subsection{Varying the stability threshold \Qmin}
\label{sec:Q}
\begin{figure}
\begin{center}
\includegraphics[scale=0.42]{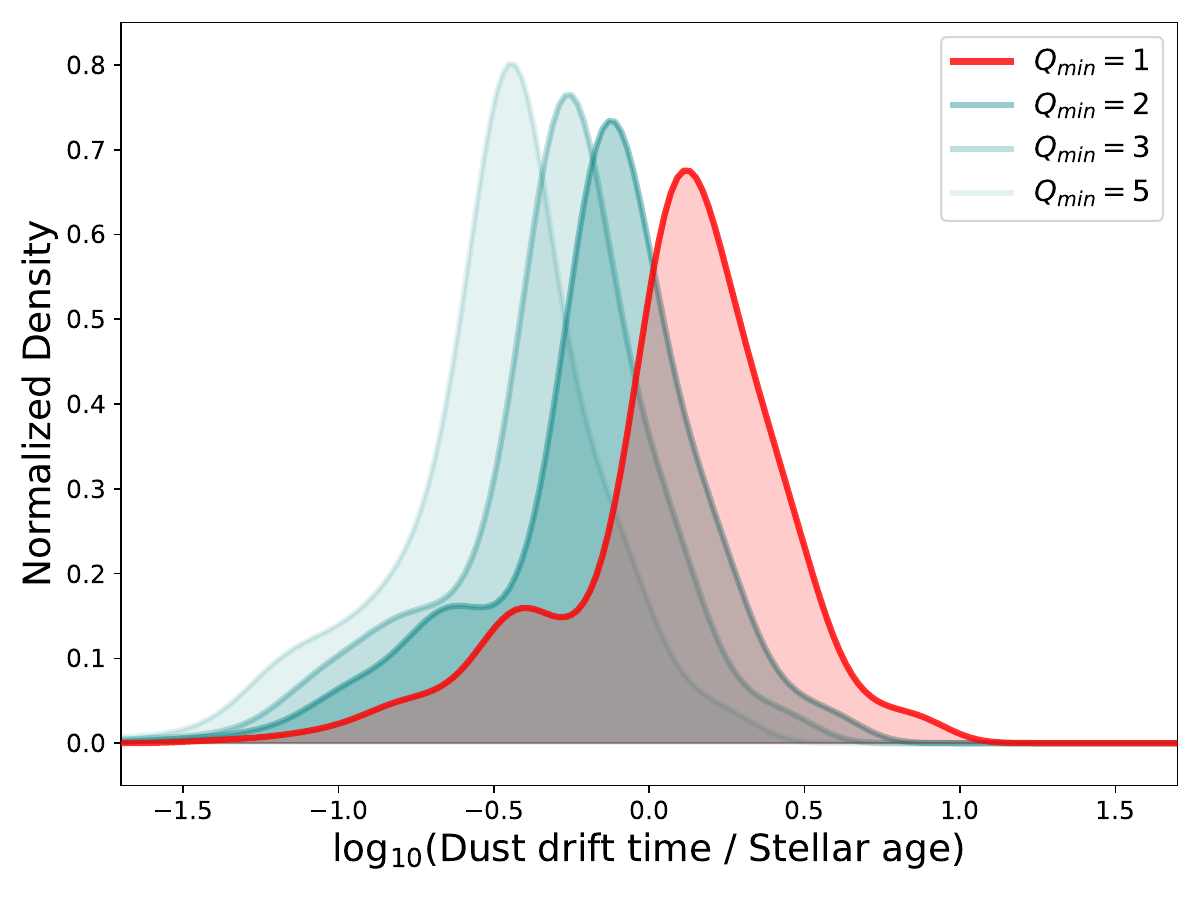}
\end{center}
\caption{The ratio of drift time to stellar age for different values of the stability threshold, \Qmin.}
\label{fig:Q}
\end{figure}

We determined the case 2 drift time by setting the minimum value of the Toomre $Q$ parameter equal to 1. However, as disks become unstable, spiral shocks and turbulence would heat the disk. Depending on the ratio of cooling time to Keplerian period, this can either lead to fragmentation or a rise in temperature, increasing the sound speed, and thereby elevating $Q$ \citep{2001ApJ...553..174G}.
Because of this tendency for disks to self-regulate at $Q > 1$, we explored the effect of a higher \Qmin\ threshold (at fixed $\gamma=1$) for the calculation of the case 2 drift time and show the results in Figure~\ref{fig:Q}.

A higher \Qmin\ threshold implies lower gas surface densities and therefore faster drift. Drift times scale approximately inversely with \Qmin\ such that the median ratios are $\log_{10}(\tdrift/t_\ast) = 0.05, -0.17, -0.32, -0.51$
for $\Qmin=1,2,3,5$ respectively. Drift times remain very long, averaging about two-thirds of the stellar age for \Qmin=2.

\subsection{Disk masses}
\label{sec:masses}
\begin{figure}
\begin{center}
\includegraphics[scale=0.42]{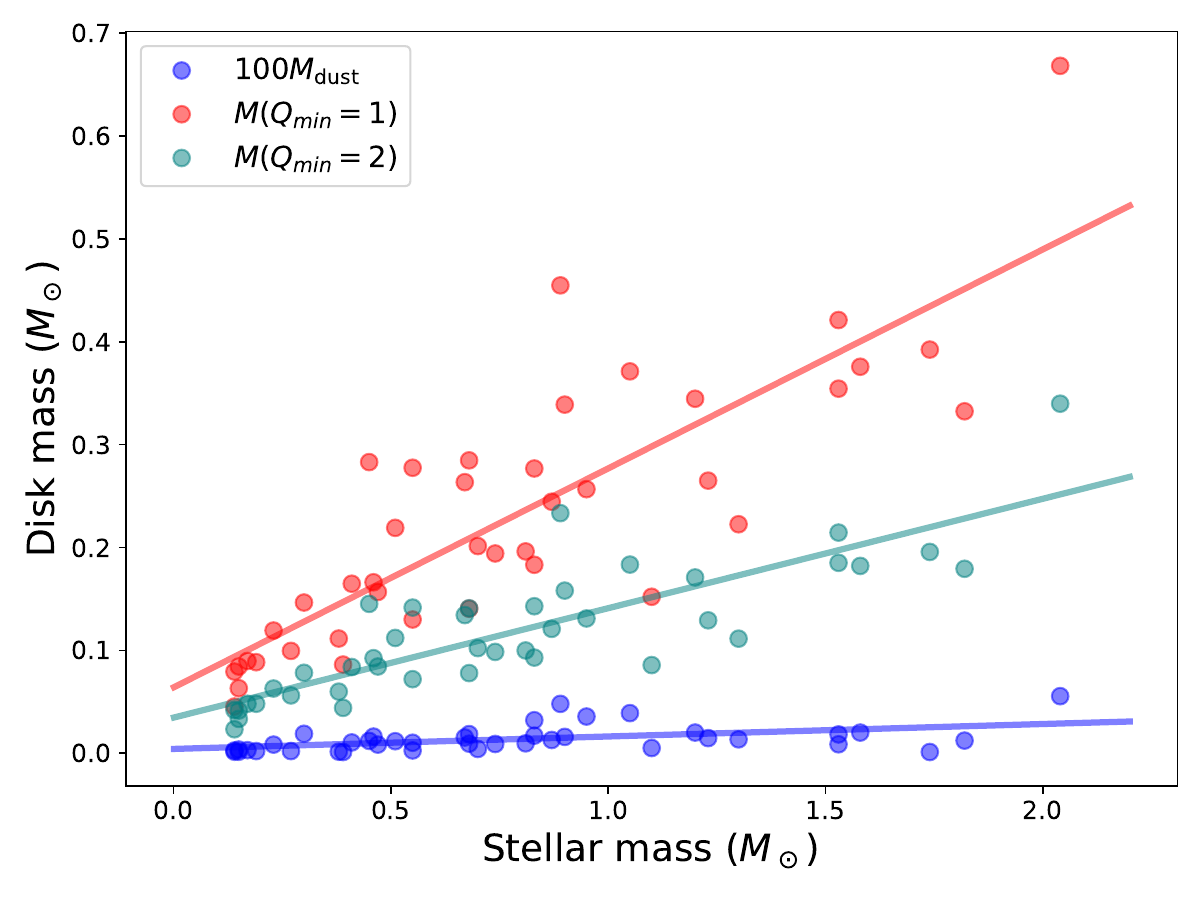}
\end{center}
\caption{The relation between disk and stellar masses for different surface density constraints. The blue points show the disk mass under the assumption that it is the dust mass multiplied by a gas-to-dust ratio of 100. The red and teal points show the masses where the minimum value of the Toomre parameter, \Qmin\ equals 1 or 2 respectively. The lines show linear fits in each case where the slope is 0.01 ($M_{\rm disk}=100M_{\rm dust}$), 0.11 ($\Qmin=2$), and 0.21 ($\Qmin=1$).}
\label{fig:mass}
\end{figure}

Disk luminosities at millimeter wavelengths are known to scale with stellar masses \citep{2013ApJ...771..129A, 2016ApJ...828...46A} and this is considered an important correlation that relates disk processes to exoplanet demographics \citep{2011A&A...526A..63A}.
We plot the disk masses for different surface density constraints in Figure~\ref{fig:mass} and overplot linear fits. For case 1, where the disk mass is equal to the dust masses scaled by a gas-to-dust ratio of 100, the blue line recaptures the known relation that $\Mdisk\simeq 0.01M_\ast$.
For case 2, where the disk mass is the maximum possible for gravitational stability, as defined as $\Qmin=1$, we find a much steeper relation, $\Mdisk\simeq 0.21M_\ast$ (red line). We also show the inferred disk masses for the slightly more relaxed constraint, $\Qmin=2$, where the drift time remains a substantial fraction of the stellar age, and find
$\Mdisk\simeq 0.11M_\ast$ (teal line).
The relation extends to higher \Qmin\ and can be approximated as $\Mdisk\simeq M_\ast/5\Qmin$. This arises because the Toomre parameter can be restated in terms of the ratio of disk to stellar mass but the scaling, which depends on the gas density distribution, is empirically determined solely from the observed CO radius.

Stellar mass accretion rates integrated over time show a wide dispersion but, on average, are somewhat closer to the case 1 disk mass constraint than case 2 \citep{2016A&A...591L...3M}. However, there is a growing sense that disk winds, whether magneto-hydrodynamic or photoevaporative, may play an understated role in disk evolution \citep{2023ASPC..534..539M}. In particular, the disks with the shortest relative drift times in Figure~\ref{fig:drift_times} have very low dust masses and a relatively small gas disk relative to its dust size as expected for photoevaporation.

The coupled gravitational stability and drift time constraints imply disk masses that range from $0.05-0.1\,M_\odot$ for low mass M dwarfs to $0.25-0.5\,M_\odot$ for Herbig A stars. This is at the limit for where HD would have been detectable with Herschel \citep{2013Natur.493..644B, 2016ApJ...831..167M} and requires very high depletion of CO and other molecules to explain their weak millimeter line emission \citep{2014ApJ...788...59W, 2017A&A...599A.113M}.
Such massive disks should show particular kinematic signatures such as ``GI wiggles'' \citep{2020ApJ...904..148H}. Very high quality data is required but such studies are underway \citep{2024ApJ...970..153A} and offer the most direct test of the case 2 hypothesis.
There are some possible signs of self-gravitational perturbations in the two most massive disks in the plot; HD\,163296 with a vortex feature on its outer ring \citep{2018ApJ...869L..49I}, and IM\,Lup which shows clear spiral features \citep{2018ApJ...869L..43H}.


\subsection{Radius evolution}
\label{sec:evolution}
\begin{figure}
\begin{center}
\includegraphics[scale=0.42]{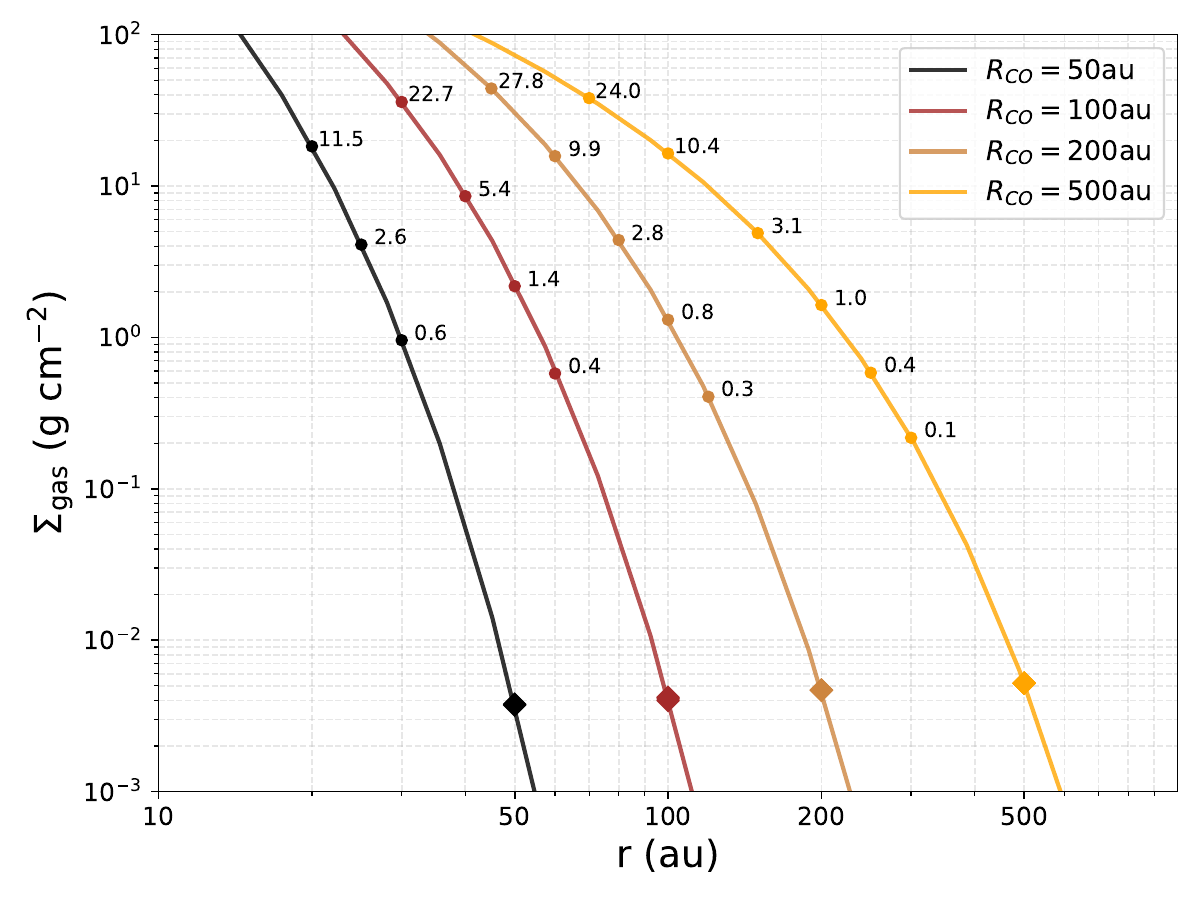}
\end{center}
\caption{Surface density profiles for a $\Qmin=1$ disk as a function of gas disk radius ranging from $\RCOninety = 50$ to 500\,au. The filled circles along each curve show the dust disk size, \Rdustninety, at $\lambda=1$\,mm for different values of the dust drift time, labeled in Myr. The profiles here all have $\gamma=1$.}
\label{fig:models}
\end{figure}

If we assume that dust drift timescales are indeed equal to the stellar age, then we can invert the reasoning here and use the gas and dust sizes as a measure of the evolutionary state of a disk. This is demonstrated in Figure~\ref{fig:models} where the dust radii is plotted for different drift times for a maximum mass gravitationally stable disk, $\Qmin=1$, at a given gas radius. The ratio of dust to gas sizes, $\Rdustninety/\RCOninety$, decreases rapidly and is well below one half within 1\,Myr, faster for large disks but slower for the most compact, dense disks, $\RCOninety=50$\,au.
The drift slows as it encounters denser gas at smaller radii with dust sizes $\sim 20-30$\% of the gas size at 10\,Myr in large disks but still $\sim 40$\% for compact disks.

These simple models do not include any evolution of the gas disk and instead offer a snapshot of where the dust line should be for a given CO radius and stellar age. They show that measured continuum sizes should always be much smaller, by at least a factor of $\sim 2$, than CO sizes in Class II sources where the stellar age $>0.5$\,Myr. Similarly, very compact dust disks that are smaller than about one fifth of the gas disk should be very rare, at least if disks masses are indeed at the cusp of gravitational instability.

\subsection{Disk substructures}
\label{sec:substructurs}
We have determined drift times using equation~\ref{eq:Powell} as if there are no pressure bumps to impede the flow of the millimeter emitting dust grains. ALMA continuum maps show that many sources are highly structured \citep{2020ARA&A..58..483A} but, based on our subjective morphological categorization at least, we do not see any relation between the presence and/or contrast of rings with drift time for either case 1 or 2. This perhaps counter-intuitive result argues against dust substructure being a solution to the drift lifetime problem although there could of course be finer scale features below detectable levels at the fidelity of the data.

The surface density fits for each source (Figure~\ref{fig:AS209}) have thicker lines where $\Sigmagas>50\,{\rm g\,cm}^{-2}$ where we expect the dust emission to become optically thick at $\lambda=1$\,mm. This shows a significant difference between the two cases, with the dust mass constraint (case 1) implying small optically thick centers, typically with radii of a few to about 10\,au, whereas the maximum mass gravitationally stable disks (case 2) should be optically thick out to several tens of au. In all cases, \Rdustninety\ lies well beyond this radius -- essentially by definition -- so the optically thick size is an extrapolation, and is fairly sensitive to the assumed value of $\gamma$, but it offers a potential test of the two cases.
Brightness temperatures can be determined from resolved observations and compared to radiative transfer models of the temperature profile to estimate the optical depth. In the DSHARP survey, the inner regions of many disks are optically thick, or nearly so, well beyond 10\,au \citep{2018ApJ...869L..42H}, especially when accounting for the effect of scattering \citep{2019ApJ...877L..18Z}.

\subsection{Broader implications}
\label{sec:implications}
The mass of a disk is a fundamental parameter in its ultimate destiny toward an exoplanetary system. What determines the mass? Stars form through the gravitational collapse of a dense, roughly solar mass, molecular core and the disk is a natural byproduct of angular momentum conservation. Given the effectively unlimited reservoir, the disk mass should settle to its maximally stable value, as defined by the Toomre parameter. How long does this situation last once the core disperses? We have shown that if the disk remains at the cusp of gravitational instability into the Class II protostellar phase, $\Qmin\sim1-2$, then the gas surface density is sufficient to prevent the drift of millimeter emitting grains at their observed radii of tens to hundreds of au.
As shown in Figure 6, these disks are $\sim10-20$ times more massive than expected based on standard conversions of the millimeter continuum flux. This implies either very high global gas-to-dust ratios $\sim (1-2)\times 10^3$ or a massive, hidden population of solids. We consider the former unlikely since the star would presumably be highly enriched in metals that is not generally seen \citep{2015A&A...582L..10K} and focus on the latter possibility. To be clear, this refers to the total mass of solids in a disk and not their spatial distribution.

We know that dust grains in disks have grown by at least three orders of magnitude from the sub-micron sizes in the ISM with an approximately power law distribution where the number of particles with size $a$ scales as $a^{-p}$ with $p\approx 3.5$ \citep{2014prpl.conf..339T}. It is likely that the size distribution does not stop there, however. Cosmochemical dating of thermally differentiated meteorites tells us that there were kilometer-sized bodies in the Solar System within $\sim 2$\,Myr of the collapse of the protosolar nebula \citep{2011AREPS..39..351D}. A likely formation mechanism is the streaming instability, which can efficiently convert about half the mass in pebbles (particles with Stokes numbers near unity) into gravitationally cohesive planetesimals on Myr timescales \citep{2016ApJ...822...55S}.

The power law size distribution does not extend to kilometers of course because the dominant physical mechanisms change with scale. However, it probably extends well beyond centimeters until fragmentation exceeds agglomeration \citep{2005A&A...434..971D}. ALMA observations therefore sample a relatively narrow range of sizes in the middle of a large distribution. For size distributions with $3<p<4$, most of the mass lies in the large particles that are not directly detected.
The striking similarity in the distribution of spectral slopes at millimeter wavelengths in different regions \citep{2021MNRAS.506.5117T} provides evidence for this interpretation. The lack of evolution in the inferred size distribution suggests that grains with radii from tens to hundreds of microns are in an equilibrium between collisional growth and destruction.

Disks are very diverse and correlations between mass, size, and stellar properties are intrinsically broad with a scatter $\sim 0.5$\,dex. Diverse evolutionary pathways should further broaden these relationships with time but this is not seen. For example, disk mass distributions have a log-normal distribution with a mean that decreases with time but an invariant dispersion \citep{2019ApJ...875L...9W}.
This is again consistent with the picture where the millimeter emitting dust is in growth-fragmentation equilibrium and relatively insensitive to any divergence at the upper end of the size distribution. The millimeter luminosity is therefore not a tracer of the total solid mass but a reflection of the quasi-static evolution in the population of larger, unseen particles.

If the mass inferred from the millimeter luminosity is indeed only $\sim5-10$\% of the total mass, the dust might be better considered as the debris from core accretion processes in the buildup of planetesimals, rather than the upper end of the size distribution due to a steady growth from ISM grains.
The continuum flux and spatial distribution would then relate to the rate and location of collisions in a planetesimal disk. The anomalously low millimeter luminosities in the young, $\sim 1$\,Myr, Ophiuchus region relative to slightly older $\sim 2-3$\,Myr Taurus and Lupus regions could then be explained as due to greater stirring of an evolving planetesimal population \citep{2019A&A...629A.116G, 2022ApJ...927L..22B}. This interpretation also more directly links stellar mass dependencies of disk and exoplanet properties.
However, the transport of volatiles from the outer disk, across snowlines, to the inner disk would be via larger particles than observed with ALMA, potentially scrambling efforts to link chemistry and disk structure.

\section{Summary}
\label{sec:summary}
We have combined recent work on protoplanetary disk sizes to measure dust drift times in a sample of 41 well resolved sources. Using an empirical relation for the gas surface density at the outer radius of the CO disk \citep{2023ApJ...954...41T}, we fit density profiles for two cases: (1) the total disk mass is the measured dust mass multiplied by a gas-to-dust ratio of 100; and (2) the disk mass is the maximum gravitationally stable mass. The inferred gas surface density at the edge of the continuum emission is directly proportional to the dust drift time \citep{2019ApJ...878..116P}. We find three main results:

\begin{itemize}
\item{Drift times are generally more than an order of magnitude shorter than the stellar age if the disk gas mass is 100 times the dust mass. This quantifies the known dust lifetime problem;}
\item{Disks that are marginally gravitationally stable, as defined through the Toomre $Q$ parameter, have sufficiently high gas densities that they can drag millimeter emitting dust grains at large radii. For $\Qmin\sim1-2$, the drift time at the observed disk continuum edge is similar to the stellar age.
This result applies over a range of ages from younger than 1\,Myr old to older than 10\,Myr and is robust to the flux threshold used to define the radius and the power law index of the surface density profile. There is no apparent correlation with disk structure in ALMA continuum images. This is an alternative solution to the lifetime problem that does not require long-lived pressure-bumps;}
\item{The CO radius determines the maximum mass that is gravitationally stable. The mass scales with the stellar mass as $\Mdisk\approx M_\ast/5\Qmin$. This is sufficient to solve the exoplanet mass budget problem.}
\end{itemize}

The similarity of dust drift times with stellar ages for disks at the cusp of gravitational instability suggests a novel interpretation of millimeter wavelength observations in protoplanetary disks whereby the continuum emission traces dust grains with high Stokes numbers from a population with a wide particle size distribution in secular equilibrium with a massive planetesimal disk.

\section*{Acknowledgments}
We thank the National Science Foundation for support through grant AST-2107841.
A.R. has been supported by the UK Science and Technology Facilities Council (STFC) via the consolidated grant ST/W000997/1 and by the European Union’s Horizon 2020 research and innovation programme under the Marie Sklodowska-Curie grant agreement No. 823823 (RISE DUSTBUSTERS project).

\vspace{5mm}
\facilities{ALMA}
\software{matplotlib \citep{Hunter:2007}, scipy \citep{2020SciPy-NMeth}. The code used to carry out the analysis and produce the figures in this paper is available at \url{github.com/interstellarmedium/dust_drift}.}

\appendix
\newpage
\section{Sample properties} \label{sec:table}
\startlongtable
\begin{deluxetable}{lcccccccccccl}
\tabletypesize{\scriptsize}
\tablewidth{0pt} 
\tablecaption{Source stellar and disk properties and computed drift times\label{tab:sample}}
\tablehead{
Source & $t_\ast$ & $L_\ast$ & $T_{\rm eff}$ & $M_\ast$ & $M_{\rm dust}$ & $R_{\rm mm, 90}$ & $\lambda$ & $R_{\rm CO,90}$ & $t_{\rm drift}^{\rm case1}$ & $t_{\rm drift}^{\rm case2}$ & $M\!(\!Q_{\rm min}\!\!=\!\!1)$ & References \\
 & (Myr) & ($L_\odot$) & (K) & ($M_\odot$) & ($M_{\oplus}$) & (au) & (mm) & (au) & (Myr) & (Myr) & ($M_\odot$) &}
\startdata 
AS209             &  1.0 &  1.41 & 4266 &  0.83 & 107 & 127 & 1.2 &  280 & 0.17 &  0.80 & 0.277 & 1 \\
CIDA1             &  1.0 &  0.20 & 3200 &  0.19 &   7 &  38 & 0.9 &  132 & 0.10 &  2.68 & 0.088 & 1 \\
CIDA7             &  2.0 &  0.08 & 3111 &  0.15 &   5 &  20 & 0.9 &   95 & 0.19 &  6.46 & 0.063 & 1 \\
CITau             &  2.5 &  0.81 & 4277 &  0.90 &  53 & 226 & 1.3 &  520 & 0.10 &  1.11 & 0.339 & 2, 3 \\
CXTau             &  1.6 &  0.25 & 3488 &  0.38 &   5 &  29 & 1.3 &  115 & 0.10 &  5.18 & 0.112 & 1 \\
CYTau             &  2.3 &  0.25 & 3560 &  0.30 &  63 &  73 & 1.3 &  295 & 0.42 &  2.75 & 0.146 & 2, 3 \\
DLTau             &  3.2 &  0.65 & 4277 &  1.05 & 130 & 145 & 1.3 &  596 & 0.73 &  5.95 & 0.371 & 1 \\
DMTau             &  3.2 &  0.14 & 3415 &  0.55 &  34 & 178 & 1.1 &  876 & 0.18 &  5.32 & 0.278 & 1 \\
DNTau             &  0.9 &  0.69 & 3806 &  0.87 &  43 &  96 & 1.3 &  241 & 0.17 &  1.71 & 0.244 & 2, 3 \\
DoAr25            &  2.0 &  0.95 & 4266 &  0.95 & 119 & 148 & 1.2 &  233 & 0.06 &  0.18 & 0.257 & 1 \\
DoAr33            &  1.6 &  1.51 & 4467 &  1.10 &  17 &  25 & 1.2 &   64 & 0.16 &  2.20 & 0.152 & 1 \\
FPTau             &  3.2 &  0.16 & 3273 &  0.39 &   4 &  47 & 1.3 &   74 & 0.01 &  0.12 & 0.086 & 1 \\
GOTau             &  2.0 &  0.21 & 3516 &  0.45 &  40 & 134 & 1.3 & 1014 & 0.15 &  4.63 & 0.283 & 1 \\
GWLup             &  2.0 &  0.33 & 3631 &  0.46 &  54 &  91 & 1.2 &  267 & 0.27 &  1.96 & 0.166 & 1 \\
HD142666          & 12.6 &  9.12 & 7586 &  1.58 &  67 &  53 & 1.2 &  172 & 0.37 &  4.05 & 0.376 & 1 \\
HD143006          &  4.0 &  3.80 & 5623 &  1.82 &  41 &  78 & 1.2 &  153 & 0.07 &  0.59 & 0.332 & 1 \\
HD163296          & 12.6 & 16.98 & 9333 &  2.04 & 185 & 137 & 1.2 &  478 & 0.48 &  3.91 & 0.668 & 1 \\
IMLup             &  0.5 &  2.57 & 4266 &  0.89 & 160 & 243 & 1.2 &  803 & 0.30 &  2.10 & 0.455 & 1 \\
IQTau             &  4.2 &  0.21 & 3690 &  0.74 &  30 & 119 & 1.3 &  220 & 0.08 &  0.64 & 0.194 & 2, 3 \\
J04202555+2700355 &  2.5 &  0.07 & 3091 &  0.14 &   5 &  34 & 0.9 &   59 & 0.02 &  0.17 & 0.045 & 1 \\
J04334465+2615005 &  1.3 &  0.12 & 3098 &  0.15 &  12 &  62 & 0.9 &  164 & 0.08 &  1.04 & 0.084 & 1 \\
J11004022–7619280 &  3.2 &  0.10 & 3270 &  0.47 &  28 & 118 & 1.3 &  273 & 0.13 &  1.36 & 0.157 & 1 \\
J16000236-4222145 &  1.3 &  0.17 & 3270 &  0.23 &  28 & 107 & 1.3 &  266 & 0.08 &  0.70 & 0.119 & 1, 4, 5 \\
J16083070-3828268 &  2.5 &  1.82 & 4900 &  1.53 &  29 &  94 & 1.3 &  394 & 0.25 &  8.63 & 0.421 & 1, 4, 5 \\
MHO6              &  2.5 &  0.06 & 3125 &  0.17 &  10 &  55 & 0.9 &  217 & 0.15 &  3.54 & 0.090 & 1 \\
MPMus             &  8.5 &  1.20 & 5000 &  1.30 &  45 &  45 & 1.3 &  110 & 0.24 &  2.04 & 0.223 & 6 \\
MYLup             &  2.0 &  0.87 & 5129 &  1.23 &  49 &  76 & 1.2 &  192 & 0.26 &  2.53 & 0.265 & 1 \\
RYLup             &  2.5 &  1.91 & 4900 &  1.53 &  61 & 121 & 1.3 &  250 & 0.11 &  0.86 & 0.354 & 1, 4, 5 \\
SR4               &  0.8 &  1.17 & 4074 &  0.68 &  31 &  29 & 1.2 &   82 & 0.26 &  2.30 & 0.141 & 1 \\
Sz100             &  1.6 &  0.08 & 3057 &  0.14 &   9 &  39 & 1.3 &  178 & 0.11 &  2.75 & 0.079 & 1, 4, 5 \\
Sz111             &  5.0 &  0.21 & 3705 &  0.51 &  39 &  85 & 1.3 &  462 & 0.35 &  6.50 & 0.219 & 1, 4, 5 \\
Sz123             &  2.0 &  0.13 & 3705 &  0.55 &   9 &  46 & 1.3 &  146 & 0.15 &  4.37 & 0.130 & 1, 4, 5 \\
Sz129             &  4.0 &  0.44 & 4074 &  0.83 &  57 &  68 & 1.2 &  130 & 0.12 &  0.64 & 0.183 & 1 \\
Sz65              &  1.3 &  0.87 & 4060 &  0.70 &  14 &  35 & 1.3 &  172 & 0.26 &  9.87 & 0.201 & 1, 4, 5 \\
Sz71              &  2.0 &  0.33 & 3632 &  0.41 &  35 &  94 & 1.3 &  218 & 0.10 &  0.83 & 0.165 & 1, 4, 7 \\
Sz84              &  1.3 &  0.12 & 3125 &  0.27 &   7 &  53 & 1.3 &  146 & 0.06 &  1.47 & 0.100 & 1, 4, 5 \\
Sz98              &  0.5 &  1.51 & 4060 &  0.67 &  51 & 163 & 1.3 &  358 & 0.07 &  0.55 & 0.264 & 1, 4, 5 \\
TWHya             &  6.3 &  0.34 & 4070 &  0.81 &  32 &  59 & 1.0 &  184 & 0.44 &  5.77 & 0.196 & 1 \\
UZTau             &  1.3 &  0.83 & 3574 &  1.20 &  67 &  84 & 1.3 &  389 & 0.70 & 10.04 & 0.345 & 1 \\
V4046Sgr          & 25.1 &  0.86 & 4350 &  1.74 &   4 &  66 & 1.1 &  362 & 0.08 & 24.30 & 0.392 & 1 \\
WaOph6            &  0.3 &  2.88 & 4169 &  0.68 &  62 &  91 & 1.2 &  298 & 0.21 &  2.15 & 0.285 & 1 \\
\enddata
\tablerefs{(1) \cite{2022ApJ...931....6L}; (2) \cite{Semenov2024PRODIGE}; (3) \cite{2018ApJ...861...64T}; (4) \cite{2018ApJ...859...21A}; (5) \cite{2021MNRAS.506.2804T}; (6) \cite{Ribas2023MPMus}; (7) \cite{2023ASPC..534..539M}.}
\end{deluxetable}

\section{Surface density fits for the full sample}
\begin{figure}[h!]
\begin{center}
\includegraphics[width=\textwidth]{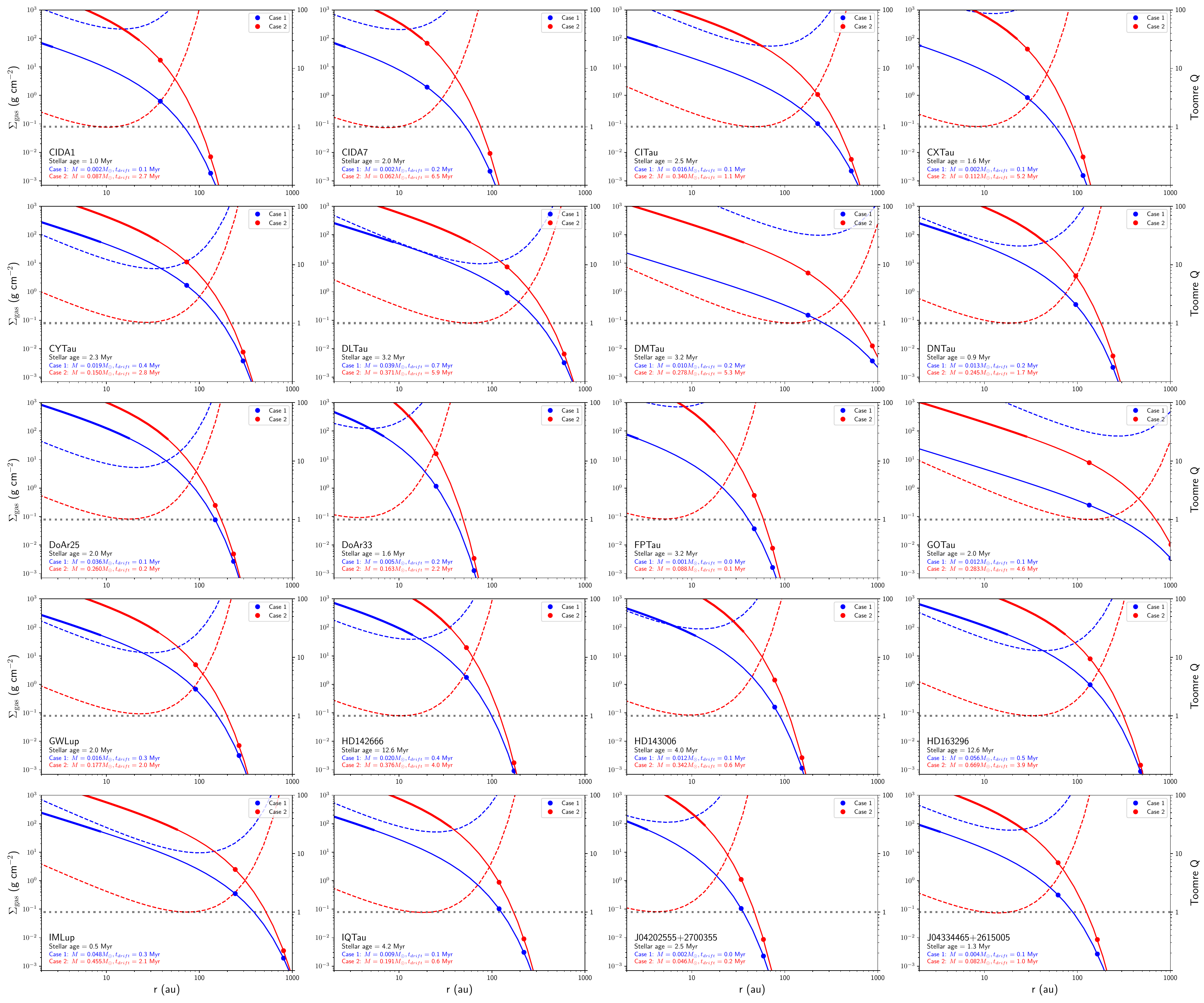}
\end{center}
\caption{The radial variation of surface density and Toomre $Q$ as in Figure~\ref{fig:AS209} but for the full disk sample (Part I).
\label{fig:panelplot1}}
\end{figure}
\begin{figure}
\begin{center}
\includegraphics[width=\textwidth]{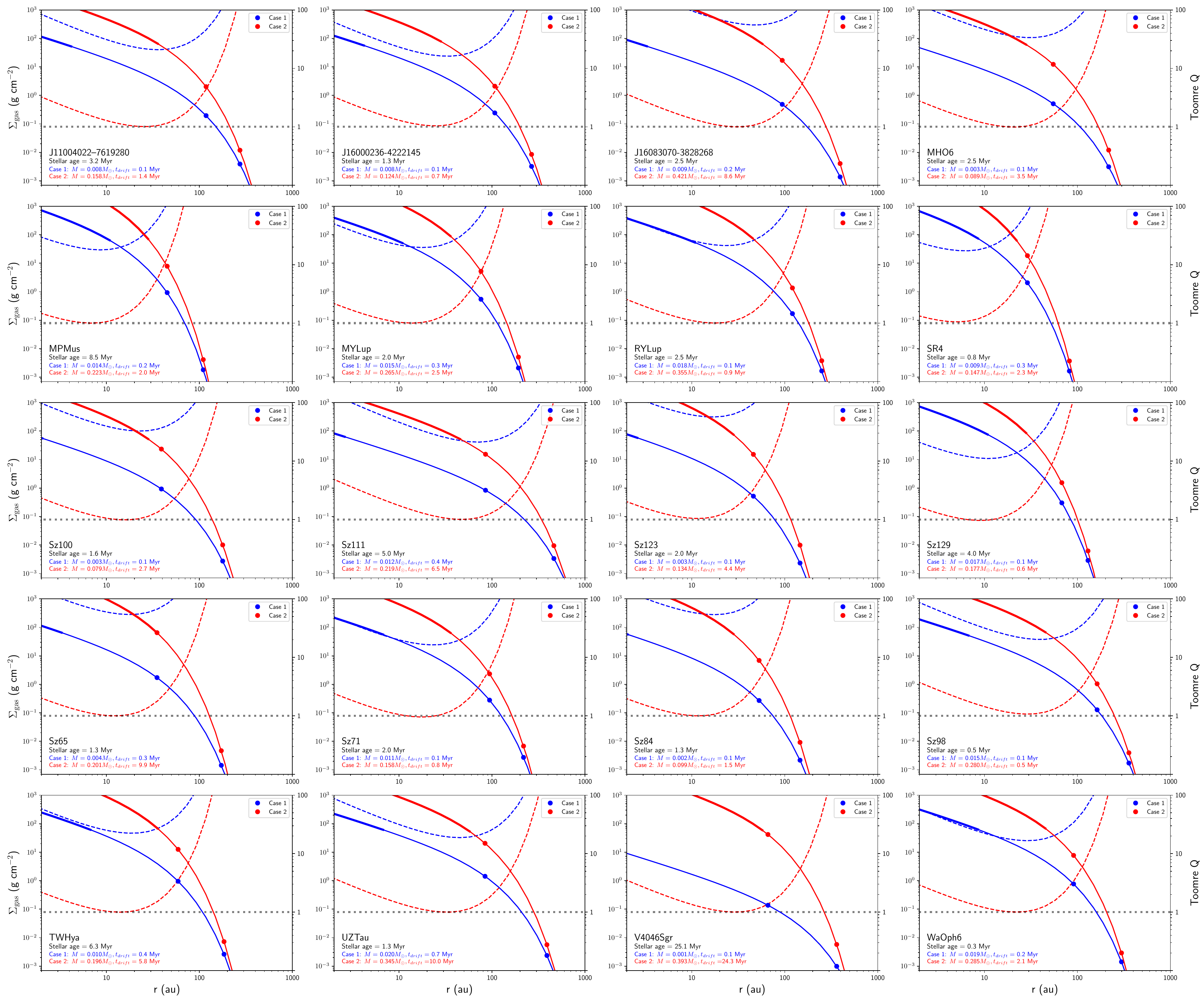}
\end{center}
\caption{The radial variation of surface density and Toomre $Q$ as in Figure~\ref{fig:AS209} but for the full disk sample (Part II).
\label{fig:panelplot2}}
\end{figure}

\newpage
\bibliography{refs, software}{}
\bibliographystyle{aasjournal}

\end{document}